\documentclass[conference]{IEEEtran} 
\usepackage{graphicx}
\usepackage{amsmath}
\usepackage{alltt}
\usepackage{times}
\usepackage{url}

\title{Inner loop optimizations in mapping single-threaded programs to hardware}

\author{Madhav P. Desai\\ Department of Electrical Engineering\\ IIT Bombay, Mumbai, India\\ madhav@ee.iitb.ac.in}

\begin{document}

\maketitle
\thispagestyle{empty}

\begin{abstract}

  In the context of mapping high-level algorithms to hardware,
  we consider the basic problem of generating an efficient hardware 
  implementation of a single threaded program, in particular, that
  of an inner loop.  
  We describe a control-flow mechanism which provides dynamic loop-pipelining
  capability in hardware, so that multiple iterations of an arbitrary inner loop 
  can be made simultaneously active in the generated hardware, 
  We study the impact of this loop-pipelining scheme in conjunction with
  source-level loop-unrolling.  
  In particular, we apply this technique to some common loop kernels: regular kernels
  such as the  fast-fourier transform and matrix multiplication, as well
  as an example of an inner loop whose body has branching.
  The resulting resulting hardware descriptions are synthesized to an FPGA target,
  and then characterized for performance and resource utilization.  
  We observe
  that the use of dynamic loop-pipelining mechanism alone typically results in
  a significant improvements in the performance of the hardware.  
  If the loop is
  statically unrolled and if loop-pipelining is applied to the unrolled
  program, then the performance improvement is still substantial.  
  When dynamic loop pipelining is used in conjunction with static loop unrolling, the improvement in performance
  ranges from 6X to 20X (in terms of number of clock cycles needed for the computation)
  across the loop kernels that we have studied.
  These optimizations do have a hardware overhead, but,
  in spite of this, we observe that the joint use of these 
  loop optimizations not only improves performance, but also the 
  performance/cost ratio of the resulting hardware.

\end{abstract}

\section{Introduction}

We consider the problem of improving the performance
of hardware generated from single threaded programs; in
particular, the important problem of mapping loops to hardware.  It
is well known that most compute intensive programs spend a large fraction
of their time in inner loops.  Thus, the optimal implementation
of such loops is of primary importance, whether the target is
a processor or hardware.  Such an improvement is essential if
synthesized hardware is to be performance competitive with
high performance processors or with hand-crafted hardware.

In the context of compilation to a pipelined processor, 
several loop optimizations
have been considered in literature, such as loop-unrolling, loop-peeling,
software-loop-pipelining etc. \cite{Wolfe}, \cite{Muchnick}, with the intent of
these optimizations being the extraction of as much
parallelism as possible from the single-threaded source program. 

Similar loop-optimization techniques have been explored and
reported in the literature related to reconfigurable hardware (for
example \cite{WeinhardtLuk}, \cite{Cardoso}, \cite{KastnerPhd}).  For instance,
in the work reported by \cite{WeinhardtLuk},
loop optimizations are done in a manner analogous to the static
techniques used in software compilers, in which index 
expressions which depend on the induction variable
are analysed to identify dependencies and schedule operations
across iterations of the loop body.  An explicitly timed controller is synthesized
for the pipeline.  Another approach which works in a similar
manner is described in \cite{Cardoso}.  These approaches rely on a static analysis of
the loop, and the cases to which the approach can be applied
are restricted (but are still sufficiently general for most linear
algebra and digital signal processing kernels).

In our approach, the hardware model is abstracted as
a virtual circuit which consists of a data-path (a graph of operations 
interconnected by wires) and a control-path
which is modeled as a Petri-net.   The operations in the data-path are
not tightly scheduled, with dependencies being taken care of
by the control Petri-net (for example: operation $X$ can start only
after operations $Y,Z$ have finished etc.). This representation allows the implementation
of loop pipelining by a simple modification to the control Petri-net
{\em without altering} the data-path.  It is possible to pipeline any loop, 
even those that do not have explicit induction variables (such as while loops).  
We will describe this loop-pipelining mechanism in a later section in
this paper.

The experimental results in this paper are based on the dynamic loop-pipelining
optimization applied by itself and in conjunction with static loop-unrolling.
By loop-unrolling, we mean a {\em static} source-level or compile-time optimization technique in
which  an inner loop is unrolled by instantiating multiple copies of the loop-body
while simultaneously reducing the number of loop-iterations.  For example:
\begin{verbatim}
for(i=0; i < 8; i++) {
   x += a[i]*b[i];
}
\end{verbatim}
is transformed to
\begin{verbatim}
for(i=0; i < 4; i+=2) {
   int i1 = i+1; int i2 = i+2; int i3 = i+3;
   x += a[i]*b[i];
   x += a[i1]*b[i1];
   x += a[i2]*b[i2];
   x += a[i3]*b[i3];
}
\end{verbatim}
This unrolling increases the size of the basic block (that is, the maximal sequence of
statements without any branches), and provides the possibility of extracting more
parallelism in the loop.  Note that this unrolling can be done manually
by the programmer, or automatically by an optimizing compiler.

In the remainder of this paper, we will first briefly describe the model of
the hardware that is produced by our HLS compiler and illustrate how this model
can incorporate dynamic run-time loop pipelining.  The chief issues here are the
hardware overhead (area, energy, delay) incurred by the need to provide 
this run-time support in the hardware generated by the compiler, and
the corresponding improvement in performance that results from this optimization.
In the results presented in this paper, the unrolling has been done 
manually at the source code level.

In order to address this issue, we will present a set of observations from
experiments performed on representative inner loops that occur in 
some important applications such as the fast-fourier transform, the
matrix product, vector dot-product, and a digital filtering algorithm.
These observations report the hardware performance on four different
loop-optimization choices: with no loop optimization, with static unrolling
alone, with dynamic loop pipelining alone, and with static unrolling combined with
dynamic loop pipelining.  Hardware resource utilization and delays are
computed by synthesizing and simulating the generated hardware for an FPGA target.

The observations indicate the following:
\begin{itemize}
\item The performance improvement with loop-pipelining applied alone is in the
2X-8X range.   This improvement is observed both in the case of an inner
loop whose body is a single basic block, as well as in the case when the
inner loop body has branches.
\item The performance improvement with loop-unrolling alone is in the 
2X range, and with aggressive unrolling (as was tried in the matrix multiplication
case), the improvement is as high as 10X.
\item The combination of loop-pipelining and loop-unrolling leads
to a performance improvement which is at least as high as the
product of the improvements due to the individual optimizations.  
For matrix multiplication, this improvement is as large as 20X.
\item The hardware overheads for implementing these optimizations are considerable,
but the cost-to-performance ratio improves substantially in all cases.
\end{itemize}
The results indicate that loop-unrolling combined with dynamic loop-pipelining
can close the performance gap noted above.


\section{A note on our compiler flow}

The dynamic pipelining mechanism described in this paper
is implemented in a compiler flow which takes 
a C program and produces an equivalent VHDL description.
We give a brief description of this compiler flow.  The
details are not relevant to this paper, and the interested
reader can find them in \cite{ahirDsd2010}.

Our compiler starts with a C program and produces VHDL.  For the
C front-end, we use the clang-2.8 compiler\footnote{www.clang.org}.
This compiler is used to emit LLVM byte-code\footnote{www.llvm.org}, 
which is then transformed to VHDL using the following transformations:
\begin{enumerate}
\item The LLVM byte-code is translated to an internal intermediate
format, which is itself a static-single assignment centric 
control-flow language (named {\bf Aa}) which allows the description of parallelism
using fork-join structures as well as arbitrary branching \cite{Aa}.
\item The {\bf Aa} description is translated to a virtual circuit (the model
is described in the next section).  During this translation, the
following major optimizations
are performed:  declared storage objects are partitioned into disjoint memory
spaces using pointer reference analysis, and dependency analysis is used to
generate appropriate sequencing of operations in order to maximize the 
parallelism.
\item The virtual circuit is then translated to VHDL.  At this point,
decisions about operator sharing are taken.  Concurrency analysis is
used to determine if a shared hardware unit needs arbitration.  Optimizations
related to clock-frequency maximization are also carried out here.
The generated
VHDL uses a pre-designed library of useful operators ranging from
multiplexors, arbiters to pipelined floating point arithmetic units.
\end{enumerate}

The compiler flow has been characterized over a wide variety
of applications \cite{ahirDsd2010}, \cite{ahirUsenix2012}.

\section{Model of the virtual circuit generated by our compiler}

The virtual circuit generated by our compiler consists of three
cooperating components: the control-path, the data-path and
the storage system \cite{ahirDsd2010}.

\begin{figure*}[ht]
  \centering
  \includegraphics[width=15cm]{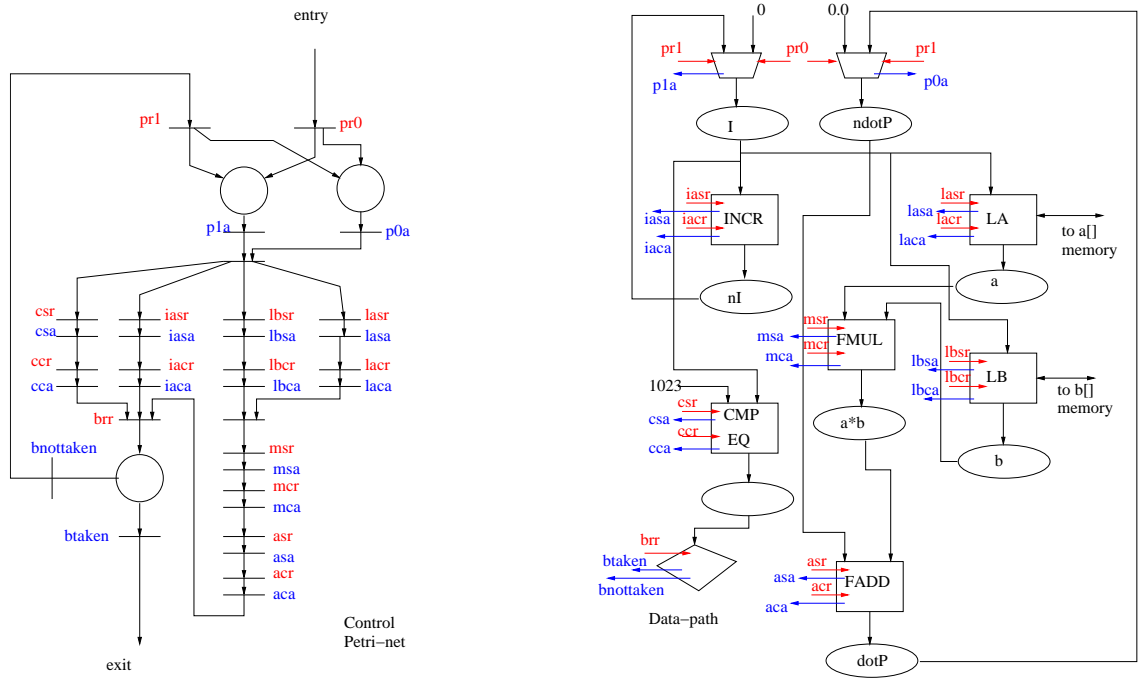}
  \caption{Control-data-storage virtual circuit model.}
  \label{fig:dotP}
\end{figure*}

To illustrate the model, we consider a simple example.
\begin{verbatim}
float a[1024], b[1024];
float dotp = 0.0;
for(i=0; i < 1024; i++)
{
   dotp += a[i]*b[i];
}
\end{verbatim}
To compile this code, we use the clang-2.8  C compiler, 
which is used to emit LLVM byte-code.  The LLVM byte-code
is transformed through a series of steps by our compiler tools to
produce a virtual circuit, which is depicted in Figure \ref{fig:dotP}.
The virtual circuit in Figure \ref{fig:dotP} has three components,
described below.

\subsection{Data-path}
The data-path is a directed hyper-graph with nodes being
operations and arcs being nets (shown as ovals).  Each
net has at most one operation which drives it.  Further, most
operations have  a split protocol handshake with
the control-path:  two pairs of request/acknowledge 
associations (*sr/*sa for sampling the inputs  and *cr/*ca for
updating the outputs).    The operation samples its inputs
on receiving the sr request symbol and acknowledges the completion
of this action by emitting the sa acknowledge symbol.  After receiving
the cr symbol, the operation will update its output net
using the newly computed value. The sequencing is required to be
\begin{verbatim}
sr -> sa -> cr -> ca
\end{verbatim}
Note that an operation can be re-triggered while an earlier
edition of the operation
is in progress (this is important if the operation is implemented
in a pipelined operator).

Some data-path operations (such as the multiplexor
shown on the top and the decision operation shown at the bottom
left in Figure \ref{fig:dotP}) follow a simpler protocol.  The multiplexor
has a pair of requests and a single acknowledge, with the condition
that at most one of the requests is received at any time instant.
The input corresponding to the request is then sampled and stored
in the output net of the multiplexor.
The decision operation has a single request and two acknowledes.  Upon
receipt of the request symbol, the decision operation checks its input net
and emits one of the two acknowledges depending on whether the input
is zero/non-zero.

In Figure \ref{fig:dotP}, the following data-path operations
are instantiated:
\begin{verbatim}
mI, mdotP  multiplexors for I, dotP.
INCR       increment for I++
LA         load for a[I]
LB         load for b[I]
FMUL       multiply for p=a[I]*b[I]
FADD       add for dotP += a*b
CMP EQ     compare for COND=(I==1023)
D          decision  COND?
\end{verbatim}

\noindent
{\bf Remark}
Note that the data-path only shows the operations and their interconnection.
When the data-path is implemented as hardware, multiple operations may
be mapped to a single operator depending on cost/performance tradeoffs.  When
this is done, multiplexing logic is introduced in the hardware.  These
decisions and manipulations are performed in the compiler stage which is
responsible for transforming the virtual circuit to VHDL.

\subsection{Storage subsystem}

The load and store operations in the data-path
are associated with memory subsystems.  In general, there
can be multiple disjoint memory subsystems inferred by our
compiler.  In this particular case, the arrays a[] and b[] 
are mapped to disjoint memories, due to which the two
loads are allowed to proceed in parallel (the relaxed consistency
model is enforced).
In order to maintain the relaxed consistency model, the
memory subsystems are designed to use a time-stamping 
scheme which guarantees first-come-first-served access to
the same memory location.

\subsection{Control-path}


The control-path in the virtual circuit encodes
all the sequencing that is necessary for correct
operation of the assembly.
The control-path (shown on the left
in Figure \ref{fig:dotP}) is modeled as a Petri-net with
a unique entry point and a unique exit point.  The Petri-net
is constructed using a set of production rules which guarantee
liveness and safeness \cite{ahirDsd2010}.  Transitions in the Petri-net
are associated with output symbols to the data-path (these can
be described by the regular expressions *sr and *cr)
and input symbols from the data-path (these are of the form *sa and
*ca).  The *sr symbols instruct an element in the data-path to
sample its inputs and the *cr symbols instruct an element in the
data-path to update its outputs (all outputs of data-path elements
are registered).  The *sa and *ca symbols are acknowledgements
from the data-path which indicate that the corresponding requests
have been served.  

The following classes of dependencies are
encoded in the control Petri-net:
\begin{itemize}
\item Read-after-write (RAW):  If the result of operator A is used
as an input to operator B, the sr symbol to B can be emitted only after
the ca symbol from A has been received.
\item Write-after-read (WAR): If B writes to a net whose value needs
to have been used by A earlier, for example as in
\begin{verbatim}
a = (b+c) -- operation A reads c
c = (p*q) -- operation B writes to c
\end{verbatim}
where there is a WAR dependency through c,
 then the cr request to B can be
issued only after the sa acknowledge from A has been received.
\item Load-Store ordering:  If P,Q are load/store operations
to the same memory subsystem, and if at least one of P,Q is a
store, and if P is supposed to happen before Q,  then the sr request
to Q must be emitted only after the sa acknowledge from Q
has been received.  The memory subsystem itself guarantees
that requests finish in the same order that they were
initiated.  This takes care of WAR, RAW and WAW memory
dependencies.
\end{itemize}

The control-path in \ref{fig:dotP} shows the sequencing
generated by these rules.  Note that the data-path
is not party to any sequencing decisions (other than
responding to the request symbols). 

\section{A control-flow mechanism for dynamic loop-pipelining}

For the subsequent discussion, we assume that the inner
loop which is being optimized consists of a single
basic block, that is, there are no branching instructions
in the loop body (no jumps, if constructs, switch constructs etc.).
If such constructs are present, these are first eliminated
using the mechanism of guarded (that is, predicated) execution.

Suppose that we want to modify the control-path in order to
permit the second (and maybe third etc.) iteration of a loop
to begin while the first iteration is still in progress.  
Consider the example of the dot product.  The original
loop was
\begin{verbatim}
float a[1024], b[1024];
float dotp = 0.0;
for(i=0; i < 1024; i++)
{
   dotp += a[i]*b[i];
}
\end{verbatim}
The {\em fully unrolled} version of this loop would be
\begin{verbatim}
float a[1024], b[1024];
float dotp = 0.0;
dotp += a[0]*b[0];
dotp += a[1]*b[1];
dotp += a[2]*b[2];
...
dotp += a[1023]*b[1023];
\end{verbatim}
Let $A$ denote the $*$ operation and let $B$ denote the $+$
operation, $L_a,\ L_b$ denote the loads from $a$ and $b$ respectively.  
In principle, all the loads can occur simultaneously, and all the multiplies
can happen simultaneously once the loads complete.  The adds
would need to be ordered because of the multiply-accumulate nature
of the code as it is written.  Any ordering of these operations which
satisfies these dependencies will be termed a {\em loop-consistent}
ordering.

\begin{figure}[ht]
  \centering
  \includegraphics[width=7cm]{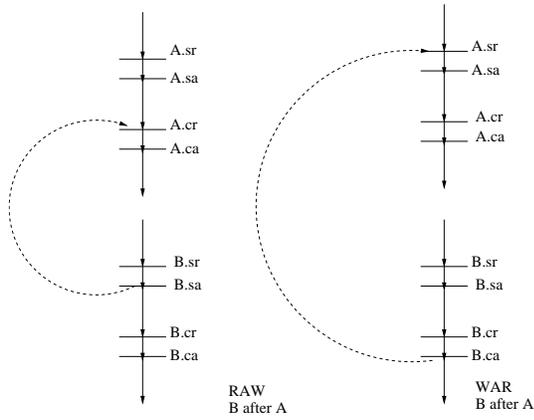}
  \caption{Control-path mechanism for handling RAW, WAR dependencies for loop-pipelining.}
  \label{fig:pipelineMechanism}
\end{figure}

In our dynamic loop pipelining scheme, we use the
following ordering scheme.
If A is an operation in the loop body,
denote the
$k^{th}$ execution of A by $A_k$.  Since each operation
has  events sr, sa, cr, ca, we denote these by
$A_k.sr$, $A_k.sa$, $A_k.cr$, $A_k.ca$ respectively. We impose 
the following dependency rules on operations across
loop iterations.
\begin{itemize}
\item $A_k.sa \rightarrow A_{k+1}.sr$ for all operations A: that is,
the next execution of A cannot start until the current execution has
finished sampling the inputs.
\item $A_k.ca \rightarrow A_{k+1}.cr$ for all operations A: that is,
the completion of the next execution of A can be initiated only
after the current execution of A has completed.
\item If $A \rightarrow B$ is a RAW dependency, then 
$B_k.sa \rightarrow A_{k+1}.cr$.  That is, until B has
sampled the result of the current A, the next completion
of A cannot start.
\item If $A \rightarrow B$ is a WAR dependency, then
$B_k.ca \rightarrow A_{k+1}.sr$.  That is, the next
A cannot start until the current B has completed.
\item If P, Q are successive load/stores, with at least
one of them being a store, then 
$Q_k.sa \rightarrow P_{k+1}.sr$.  That is, the next P
cannot start until the current Q has acknowledged that it
has started.
\end{itemize}
The mechanism for incorporating RAW and WAR dependencies is
illustrated in Figure \ref{fig:pipelineMechanism}
for RAW and WAR dependencies within the loop body.
The reverse dotted arc is a marked arc (it initially
carries a single token).

\begin{figure}[ht]
  \centering
  \includegraphics[width=7cm]{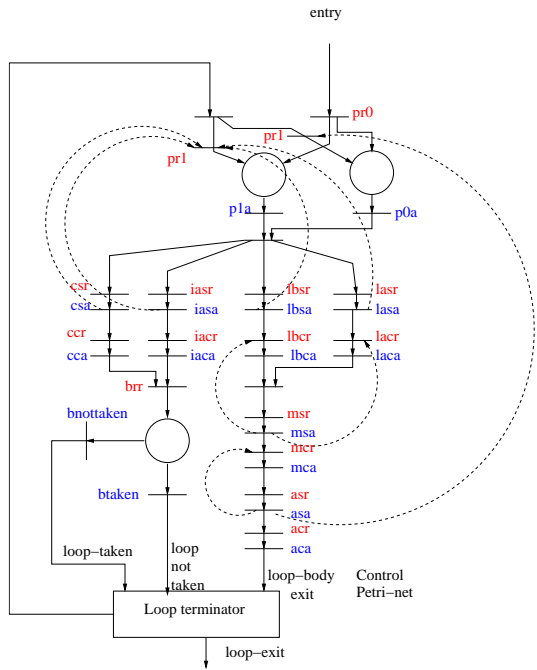}
  \caption{Modified control-path with loop-pipeline dependencies (dotted lines).}
  \label{fig:pipelinedCP}
\end{figure}

It is easy to confirm that these additional dependencies ensure
that the loop execution subject to these dependencies is a
loop-consistent ordering.  
The modified control path is shown in Figure \ref{fig:pipelinedCP}.
The loop-terminator element has three inputs:  a loop-taken transition,
a loop-not-taken transition and a loop-body-exit transition.  The loop-taken/not-taken
pair indicates whether a new iteration is to be started or whether the 
loop has terminated.  The loop-body-exit
transition indicates that the body of the loop has finished executing an
iteration.  The loop-terminator initiates a new iteration as long as the
number of active iterations is within a specified limit $M$ (usually, we keep
this limit to $M=8$).  Thus, all the places in the modified control path
in Figure \ref{fig:pipelinedCP} now must have a capacity of $M$ and the
cost of implementing each place in the control path goes up by a
factor of $\log M$.  This is the major additional cost incurred by
the pipelining mechanism.

\section{Experimental Results}

We considered four examples:
\begin{itemize}
\item Three examples where the loop body was a single basic block: the dot product, the fast-fourier-transform,
and matrix multiplications, each with critical inner loops.  Four 
configurations were tested in each case: the basic code, the unrolled code,
the basic code with loop-pipelining and the unrolled code with loop-pipelining.
In each case, the generated VHDL code was synthesized to a Xilinx Virtex-6
FPGA and synthesis results were used to estimate the resource usage,
the clock frequency and the number of cycles required by the inner
loop.
\item One example where the loop body exhibits branching: a stream processor kernel which
operates on a stream of numbers and performs operations on the stream depending on
an op-code stream.  This example illustrates that the dynamic loop-pipelining mechanism
is effective in complex loop bodies as well.
\end{itemize}

\section{The dot product}

The basic code (a,b are arrays, dotP is the accumulated dot-product):
\begin{verbatim}
for(I=0; I < 64; I++)
{
  dotP += a[I]*b[I];
}
\end{verbatim}

The unrolled version of the code used:
\begin{verbatim}
for(I=0; I < 64; I += 4)
{
  I1 = I+1; I2 = I+2; I3 =I+3;
  dotP += ((a[I]*b[I]) + (a[I1]*b[I1]) 
           + (a[I2]*b[I2]) + (a[I3]*b[I3]));
}
\end{verbatim}

The results are shown in Table \ref{table:dotP}.   The number of clock-cycles
needed to complete the inner loop, the number of look-up tables needed, the
 number of flip-flops needed and the post-synthesis clock frequency estimate
are reported in the table.  The last two rows correspond to the normalized
performance (time needed by the plain case relative to the optimized case, higher is better),
and the normalized performance/cost ratio (time/(LUTs+FF)  ratio normalized
with respect to the plain case, higher is better).
\begin{table}[htb]
  \centering
  \caption{Dot-product results with and without loop-optimizations}
  \label{table:dotP}
  \renewcommand\arraystretch{1.2}
  \setlength{\tabcolsep}{1ex}
  \begin{tabular}{c|c|c|c|c}
  \hline
  & plain & pipelined & unrolled & pipelined  \\
  &       &           &          & + unrolled\\
  \hline

Cycles   &    4071  &   1874  &     1894  &     582 \\
LUTs     &   4288   &   5195  &     4809  &    7344 \\
FFs      &    3799  &   4345  &     4378  &    5911 \\
Freq.(MHz)  & 199.7  &    199.7 &      199.7 &    199.7 \\
Norm. Perf. & 1      &  2.17  & 2.15 & 7 \\
Norm. Perf/Cost & 1  &  1.83  & 1.89 & 4.27 \\
\hline
  \end{tabular}
\end{table}

From Table \ref{table:dotP}, we see 
\begin{itemize}
\item If loop-pipelining is 
applied to the plain program, performance improves
by about 2X relative to the non-pipelined, plain case.
\item If loop-pipelining is applied to the unrolled program,
performance improves by more than 3X relative to the non-pipelined,
unrolled case.  
\item In terms of the performance/cost
ratio, the pipelined-unrolled version is more than 3X better than
the plain version.   The normalized performance/cost ratio is 4X
better when both loop optimizations are used.
\end{itemize}

\section{The fast-fourier-transform (FFT)}

A 64 point FFT program (radix two, in-place, twiddle factors computed
apriori) with the following loop-structure was used:
\begin{verbatim}
for(STAGE=0; STAGE < 6; STAGE++)
{
  for(BFLY=0; BFLY < 32; BFLY++)
  {
    Butterfly(STAGE,BFLY);
  }
}
\end{verbatim} 
where the function {\bf Butterfly} is inlined and implements
the radix-two butterfly with index BFLY in stage STAGE using
the precomputed twiddle factors.
In the unrolled version, the inner-loop was rewritten as
\begin{verbatim}
  for(BFLY=0; BFLY < 32; BFLY += 4)
  {
    Butterfly(STAGE,BFLY);
    Butterfly(STAGE,BFLY+1);
    Butterfly(STAGE,BFLY+2);
    Butterfly(STAGE,BFLY+3);
  }
\end{verbatim}

The results are shown in Table \ref{table:fft} (the cycle count is for a single
stage of the FFT).
\begin{table}[htb]
  \centering
  \caption{FFT-results  with and without loop-optimizations}
  \label{table:fft}
  \renewcommand\arraystretch{1.2}
  \setlength{\tabcolsep}{1ex}
  \begin{tabular}{c|c|c|c|c}
  \hline
  & plain & pipelined & unrolled & pipelined  \\
  &       &           &          & + unrolled\\
  \hline

Cycles   &    4151  &   2885   &     3110  &     1064 \\
LUTs     &   12155   &  23139  &     16032  &    37831 \\
FFs      &    11955  &  18632  &     15299  &    28698 \\
Freq.(MHz)  & 186.9  &    164.1 &      186.9 &    164.7 \\
Norm. Perf. & 1      &  1.26  & 1.33 & 3.42 \\
Norm. Perf/Cost & 1  &  0.73  & 1.02 & 1.24 \\
\hline
  \end{tabular}
\end{table}
From Table \ref{table:fft}, we see 
\begin{itemize}
\item If loop-pipelining is 
applied to the plain program, performance improves
by about 1.26X relative to the non-pipelined, plain case.
\item If loop-pipelining is applied to the unrolled program,
performance improves by more than 2.5X relative to the non-pipelined,
unrolled case.  
\item In terms of the performance/cost
ratio, the pipelined-unrolled version is only 1.24X better than
the plain version.   
\end{itemize}
The reason for the poorer results in this case is the use of the in-place
algorithm.  The bottleneck in this case becomes the access to the memory
subsystem in which the array is stored. 

\section{Matrix multiplication}

The plain triple loop matrix multiplication algorithm was
used as a starting point.
\begin{verbatim}
float a[16]16], b[16][16], c[16][16];
for(i = 0; i < 16; i++) {
  for(j=0; j < 16; j++) {
    float v = 0.0;
    for(k = 0; k < 16; k++) {
       v += a[i][k]*b[k][j];
    }
    c[i][j] = v;
  }
}
\end{verbatim}
The unrolled version was aggressively generated so that
the inner loop simultaneously computes 16 entries of the
product at a time.
\begin{verbatim}
float a[16]16], b[16][16], c[16][16];
for(i = 0; i < 16; i += 4) {
  for(j=0; j < 16; j += 4) {
    float v00 = 0.0, v01 = 0.0,
          v02 = 0.0, ... v33 = 0.0;
    for(k = 0; k < 16; k += 4) {
       v00 += (a[i][k]*b[k][j] 
             + a[i][k+1]*b[k+1][j]
             + a[i][k+2]*b[k+2][j]
             + a[i][k+3]*b[k+3][j]);
       ...
       v33 += (a[i+3][k]*b[k][j+3] 
             + a[i+3][k+1]*b[k+1][j+3]
             + a[i+3][k+2]*b[k+2][j+3]
             + a[i+3][k+3]*b[k+3][j+3]);
    }
    c[i][j]   = v00;
    ..
    c[i+3][j+3] = v33;
  }
}
\end{verbatim}

The observations are shown in Table \ref{table:mmultiply}.
\begin{table}[htb]
  \centering
  \caption{Matrix-multiplication-results  with and without loop-optimizations}
  \label{table:mmultiply}
  \renewcommand\arraystretch{1.2}
  \setlength{\tabcolsep}{1ex}
  \begin{tabular}{c|c|c|c|c}
  \hline
  & plain & pipelined & unrolled & pipelined  \\
  &       &           &          & + unrolled\\
  \hline

Cycles   &    161K  &   77K   &     13K  &     7810 \\
LUTs     &   6323   &  9408  &     14891  &    31744 \\
FFs      &    6974  &  8818  &     12041  &    21060 \\
Freq.(MHz)  & 199.7  &   199.7 &   186.1 &    164.1 \\
Norm. Perf. & 1      &  2.09  & 11.5 & 16.9 \\
Norm. Perf/Cost & 1  &  1.52  & 5.67 & 4.25 \\
\hline
  \end{tabular}
\end{table}
From Table \ref{table:fft}, we see 
\begin{itemize}
\item If loop-pipelining is 
applied to the plain program, performance improves
by about 2X relative to the non-pipelined, plain case.
\item If loop-pipelining is applied to the unrolled program,
performance improves by 1.5X relative to the non-pipelined,
unrolled case.  
\item In terms of the performance/cost
ratio, the pipelined-unrolled version is only 4.25X better than
the plain version.   
\end{itemize}
In this instance, the aggressive loop-unrolling shows
excellent performance.  Loop-pipelining when combined with loop-unrolling,
gives a 20X improvement in the cycle count.  The normalized
performance and performance/cost improvements are also substantial.

\section{A stream processor}

The following loop was used to test a situation in
which the loop body has branching.  
\begin{verbatim}
while(1)
{
  float x = read_float32("x_pipe");
  float y = read_float32("y_pipe");
  uint8_t op_code = read_uint8("op_pipe");

  float result = 0;
  if(op_code == 0)
    result = x*y;
  else if(op_code == 1)
    result = x+y;
  else if(op_code == 2)
    result = (x*x) - (y*y);
  else if(op_code == 3)
    result = (x + y) * (x + y);
  else
    result = 0;
  write_float32("z_pipe",result);
}
\end{verbatim}
In this loop, $x,y$ and $op\_code$ are read from
three input streams.  Depending on the value of
$op\_code$, a result is calculated and written out
to an output stream.

In order to pipeline this loop, the conditional
statements are first eliminated using guards.   This
is done by calculating the predicates 
\begin{verbatim}
(op_code == 0)
(op_code == 1)
(op_code == 2)
(op_code == 3)
\end{verbatim}
and using these predicates to guard the execution of
the statements which depend on these conditions.  This
is done automatically in our compiler.

The observations are shown in Table \ref{table:streamProcessor}.
The time reported is that needed to process 16 elements from
the streams (that is, to complete 16 iterations of the loop).
\begin{table}[htb]
  \centering
  \caption{Stream processor inner loop observations: pipelined versus non-pipelined}
  \label{table:streamProcessor}
  \renewcommand\arraystretch{1.2}
  \setlength{\tabcolsep}{1ex}
  \begin{tabular}{c|c|c}
  \hline
  & plain & pipelined \\
  \hline

Cycles   &    2913  &   334  \\
LUTs     &    4603  &  12501  \\
FFs      &    4248  &  8808  \\
Freq.(MHz)  & 184.9  &   103.9 \\
Norm. Perf. & 1      &  4.86 \\
Norm. Perf/Cost & 1  &  2.01  \\
\hline
  \end{tabular}
\end{table}

We observe a 4X improvement in performance and a 2X improvement
in the performance/cost ratio.

\section{Conclusion}

We have considered the problem of optimizing inner loop
implementations in an algorithm-to-hardware compilation system.
Two optimizations were considered: static source-level loop
unrolling and dynamic hardware supported loop-pipelining.
The loop-pipelining mechanism is implemented by modifying
the control-flow in the generated hardware (without disturbing
the data-path).

The data obtained from four inner loop
kernels is encouraging. The first three  were examples in
which the inner loop body consisted of a single basic 
block. In these cases, both loop-pipelining
and loop-unrolling lead to substantial performance gains.  Further,
using both optimizations together results in multiplicative gains
and in call cases, leads to hardware which is substantially
faster and more efficient (in terms of the performance/cost ratio).
The performance gain is lower if there is a bottleneck in the
algorithm itself, such as in the FFT case, in which accesses to
the in-place array reduce the performance gains seen due to
the loop optimizations.
In the fourth inner loop kernel, we considered a loop body
which had branching.  In this case, considerable performance
and performance/cost gains were observed when loop pipelining
was enabled.

Thus, the use of hardware based dynamic loop-pipelining techniques 
offers a significant boost in performance in 
hardware synthesized from single-threaded programs.  
The performance boost provided by the dynamic loop-pipelining
in hardware seems to indicate that its use,
especially in conjunction with aggressive loop unrolling can offer 
a substantial reduction in the gap between 
the quality of automatically generated hardware and hand crafted 
hardware implementations of the same algorithm.  This needs
to be investigated further.

\bibliography{ref}
\bibliographystyle{IEEEtran}

\end{document}